\documentclass[12pt]{article}
\usepackage[a4paper]{geometry}
\usepackage{enumerate}
\usepackage{mathtools}
\usepackage{amsmath}
\usepackage{graphicx}
\usepackage{epsfig}
\usepackage{sidecap}
\usepackage{enumerate}
\usepackage{hyperref,stackrel}
\usepackage[normalem]{ulem}
\usepackage[bottom]{footmisc}
\usepackage{color}
\usepackage{enumerate}
\usepackage{bm}
\usepackage[utf8]{inputenc}
\usepackage{amsmath}
\usepackage{amsfonts}
\usepackage{amssymb}
\usepackage{graphicx}
\usepackage{enumitem}
\usepackage{authblk}
\usepackage{hyperref}
\usepackage [english]{babel}
\usepackage{braket}
\usepackage{color}
\usepackage{xcolor}
\usepackage{longtable}
\usepackage[autostyle]{csquotes}
\MakeOuterQuote{"}

\author[1]{Abhishek Yadav\thanks{2019007@iiitdmj.ac.in}}
\author[2]{Sandeep Mishra\thanks{sandeep.mtec@gmail.com}}
\author[2,@]{ Anirban Pathak\thanks{anirban.pathak@gmail.com}}
\affil[1]{PDPM Indian Institute of Information Technology Design and Manufacturing Jabalpur, India 482005}
\affil[2]{Jaypee Institute of Information Technology, A 10, Sector 62, Noida,
UP-201309, India}
\affil[@]{Corresponding author:anirban.pathak@gmail.com}

\title{Partial Loopholes Free Device Independent Quantum Random Number Generator Using IBM's Quantum Computers}
 
\begin{document}

\maketitle

\begin{abstract}
    Random numbers form an intrinsic part of modern day computing with applications in a wide variety of fields. But due to their limitations, the use of pseudo random number generators (PRNGs) is certainly not desirable for sensitive applications. Quantum systems due to their intrinsic randomness form a suitable candidate for generation of true random numbers that can also be certified. In this work, the violation of CHSH inequality has been used to  propose a scheme by which one can generate device independent quantum random numbers by use of IBM quantum computers that are available on the cloud. The generated random numbers have been tested for their source of origin through experiments based on the testing of CHSH inequality through available IBM quantum computers. The performance of each quantum computer against the CHSH test has been plotted and characterized. Further, efforts have been made to close as many loopholes as possible to produce device independent quantum random number generators. This study will help provide new directions for the development of self-testing and semi-self-testing random number generators using quantum computers.
\end{abstract}


\section{Introduction}
Randomness \cite{rand} is an essential part of human life as one knowingly or unknowingly observes/uses some random phenomena or becomes a source for producing random numbers. In fact, as of now, random numbers have become an inseparable part of modern computing with applications in a wide range of fields, including but not restricted to cryptography \cite{RNGcryp,gennaro2006randomness}, simulation\cite{RNGsimu}, gaming\cite{magerkurth2004towards,helmer1972cross}, Monte-Carlo simulation \cite{Monte1,Monte2,Monte3} and weather prediction \cite{buizza2001chaos,palmer2005representing,buizza1999stochastic}. Although random numbers are  used in almost all domains of science, most of the applications do not require perfect randomness. However, casino owners, users of cryptography including common citizens performing e-banking and many others ideally require perfect random numbers.  In reality, the perfectness requirement is not strictly implemented, but the quality of the randomness produced by a random number generator is checked through a set of tests. If one wishes  to use 
some random numbers for cryptography, then those numbers have to necessarily pass a set of tests like NIST tests \cite{NIST} and diehard test \cite{diehard}. There are many ways to produce random numbers. Before we discuss various methods that can be used to produce randomness, we may note that based on the source and method of extraction of random numbers, we can broadly classify the existing random number generators into two classes: 
\begin{itemize}
    \item{Pseudo random number generators (PRNGs)}: The  easiest method of producing random numbers is via the use of some mathematical functions. Such type of random number generators is known as PRNGs \cite{prng,prngFPGA,prngOSIH,prngLPA}. In fact,  PRNGs are based on some deterministic algorithms that use a small seed (random number) with some minimum entropy to generate a larger sequence of random numbers. The generated sequence imitates the distribution of random numbers, but in reality, such a sequence will have a periodicity. Moreover, the random numbers generated from such a process can become vulnerable if an adversary gets hold of the seed value. The advantage associated with PRNGs is that they are very fast. As of now, PRNG such as Blum Blum Shum generator (BBS) is considered to be cryptographically secure (CSPRNG\cite{blum1986simple}) and is used in many applications.
    
    \item{ True random number generators (TRNGs)}: TRNGs use some physical system that can act as a source of randomness such as the decay of a radioactive element \cite{radioactTrng}, current or voltage fluctuation  \cite{TRNG1}, or some complex phenomena such as time of arrival of cosmic photons \cite{PhysRevLett.118.060401}. The complex nature of these physical systems and the associated uncertainties leading to unpredictability makes them a good source of randomness, but such systems are not entirely free of limitations. These physical systems are too complex, and consequently, we cannot certify the amount of randomness present in them\cite{trnglast}. Further, the outcomes of these physical sources which seem to behave randomly to us as of now, may become predictable with the advent of a more mature theory. Of all the existing TRNGs, the quantum random number generator (QRNG) forms the most important and special subclass. The intrinsic randomness (probabilistic nature) of quantum mechanics is used in QRNGs to generate random numbers of very high quality \cite{radioactTrng,mannalath2022comprehensive}. Further, the random numbers generated by such generators are private too. QRNGs are  based on the fundamental phenomena of quantum physics such as superposition, entanglement, collapse on measurement to generate random numbers.  Moreover, the source of randomness used in QRNGs  is such that their entropy can be certified\cite{mannalath2022comprehensive}. As of now, several  QRNGs based on optical and non-optical systems have been designed and experimentally demonstrated to produce random numbers of very high quality. In fact, several off-the-shelf QRNGs are currently available commercially \cite{qnulabs, toshiba_qrng, quintessence_qrng, id_quantique_rng}. 
    \end{itemize}
Even though many QRNGs are available commercially, but these QRNGs are trusted devices. While using such a device, we have to trust the manufacturer or equivalently the device that the source of randomness is actually  quantum mechanical in nature. So, there is a lack of proper mechanisms which can be used to certify the source of randomness. In fact, another worrisome issue can arise if the device is under the control of an adversary. The solution lies in utilizing the nonlocal properties of quantum states to generate random numbers that are truly random as well as private. The idea is to certify the source of randomness by use of tests such as Bell tests, CHSH test \cite{bell1964einstein,clauser1969proposed}.  The devices that use the nonlocal properties of quantum states to generate certifiable random numbers form a new subclass of TRNGs which is referred to as device independent QRNGs (DI-QRNGs) \cite{pironio2010random}.  Currently the researchers world over are looking for ways to develop DI-QRNGs as their output is of very high quality and these devices can be tested against an attack by an adversary \cite{pironio2010random,giustina2013bell,liu2018device,liu2021device}. If we look through the technological advancements, quantum computers have now become a reality even though the number of qubits are not so high \cite{arute2019quantum,cross2018ibm,monroe2021ionq,macquarrie2020emerging}. Many countries government as well as big corporations such as IBM and Google are investing heavily on quantum computers. In fact, IBM has put their quantum computers on the cloud for access. Many researchers have tried to utilize the cloud based access of quantum computers to generate quantum random numbers of high quality \cite{tamura2020quantum,tamura2021quantum,li2021quantum,jacak2020quantum,kumar2022quantum,bhatia2022generation,salehi2022hybrid}. In 2020,  Tamura and Shikano used "IBM 20Q Tokyo" quantum computer to apply Hadamard gate on 20 qubits and then noted down measurement results in computational basis. They performed the experiment for more than one month by concatenating the output and finally obtained random number of length 43560 bits \cite{tamura2020quantum}.  Subsequently, the process was repeated using  "IBM 20Q
Poughkeepsie" quantum computer with  8192 shots to obtain a binary sequence with a length of 8192 per qubit \cite{tamura2021quantum}. The experiment was repeated many times to obtain a statistically large set of random numbers which was subjected to NIST tests. Similar proposals using IBM quantum computers were suggested by Kumar et al. \cite{kumar2022quantum}, Bhatia et al. \cite{bhatia2022generation}. In 2022, Salehi et al. proposed  a hybrid protocol of using Hadamard and controlled-Hadamard gates in IBM quantum computers to generate random numbers \cite{salehi2022hybrid}. In 2020, Jacak et al. proposed a scheme in which random numbers are generated by measurement upon entangled states  that can be implemented using cloud based quantum computers \cite{jacak2020quantum}. In 2021, Li et al. proposed a scheme by incorporating estimation of the errors in the preparation for superposition states to generate certifiable random numbers using IBM quantum computers \cite{li2021quantum}. However, to the best of our knowledge, as of now no effort has been made to study device independent quantum random numbers using the IBM quantum computers in particular and the quantum computers available over the cloud in general. Motivated by these facts, here we plan to investigate the possibilities of utilizing IBM quantum computers available over the cloud as DI-QRNGS.

The rest of the paper is organized as follows. In the next section \ref{qrng_basics}, we will first introduce how cloud based quantum computers in general and IBM quantum computers in particular can be used to generate quantum random numbers. Further, we will introduce the notion of self-testing  and semi-self-testing QRNGs. Following that in section \ref{CHSH_QRNG}, we will explain how certifiable quantum random numbers can be generated using violation of CHSH inequality. Further, the scheme will be implemented using different available IBM quantum computers. Then in section \ref{loopholes}, we will try to close the loopholes so that the scheme can be used to generate certifiable random numbers. Finally, we conclude our results in section \ref{conclusions}.   

\section{Can we use quantum computers as QRNGs or self-testing QRNGs?} \label{qrng_basics}
Yes, quantum computers can be used to generate random numbers. Though it's true for quantum computers in general, we will restrict our discussion to IBM's quantum computers as we have used them for this work. Before we describe the ways in which IBM quantum computers can be used as QRNGs, we may note that they are based on superconducting transmon qubits located in dilution refrigerators. A user located anywhere in the world can create quantum circuits using a  python library called 'qiskit' \cite{qiskitM} or other methods and run the circuits on any of the quantum computers available on cloud. Such a quantum computer available over the cloud can be used for performing various computational tasks including the generation of random numbers. To generate random numbers using a quantum computer over cloud, we primarily need to design appropriate quantum circuits while  keeping the size constraints of the available quantum computers in mind. In what follows, we will first elaborate on  two specific methods by which one can generate the random numbers using quantum computers and then we will address the need for self-testing and semi-self-testing QRNGs. Let us start with the simplest approach of quantum random number generation using Hadamard gate(s). 

\subsection{QRNG based using Hadamard gate(s)}

We all know that, when a single photon passes through a beamsplitter, then it is simultaneously  present in both the transmitted as well as reflected path. But if we put detectors on both the paths, then the photon will be detected in only one of the paths. This is because, before the measurement, the photon can be simultaneously present in both reflected and transmitted paths, but when detectors are placed then due to the collapse on measurement principle the photon is randomly detected by either of the detectors. We may like to note that Quantis random number generator works on this principle \cite{id_quantique_rng}. In fact, the beam splitter acts like a Hadamard gate as it creates an equal superposition of two orthogonal states (say, an equal superposition of two states $\ket{0}$ and $\ket{1}$ states, where $\ket{0}$ and $\ket{1}$ can be orthogonal states of photon, transmon qubits or any other realization of qubits). Now, the above described optical realization of random number generation can be mimicked in any quantum computer (including IBM quantum computers) as one or more Hadamard gates can be used to create superposition of orthogonal states and subsequent measurement can randomly collapse the superposition state into one of the possible states. Multiple runs of such measurements may lead to a long string of random numbers by concatenating the outcome of each measurement with the outcome of the previous measurements. Let us elaborate this idea with an example.

\begin{figure}[h!]
    \centering
    \includegraphics[width = .8\textwidth]{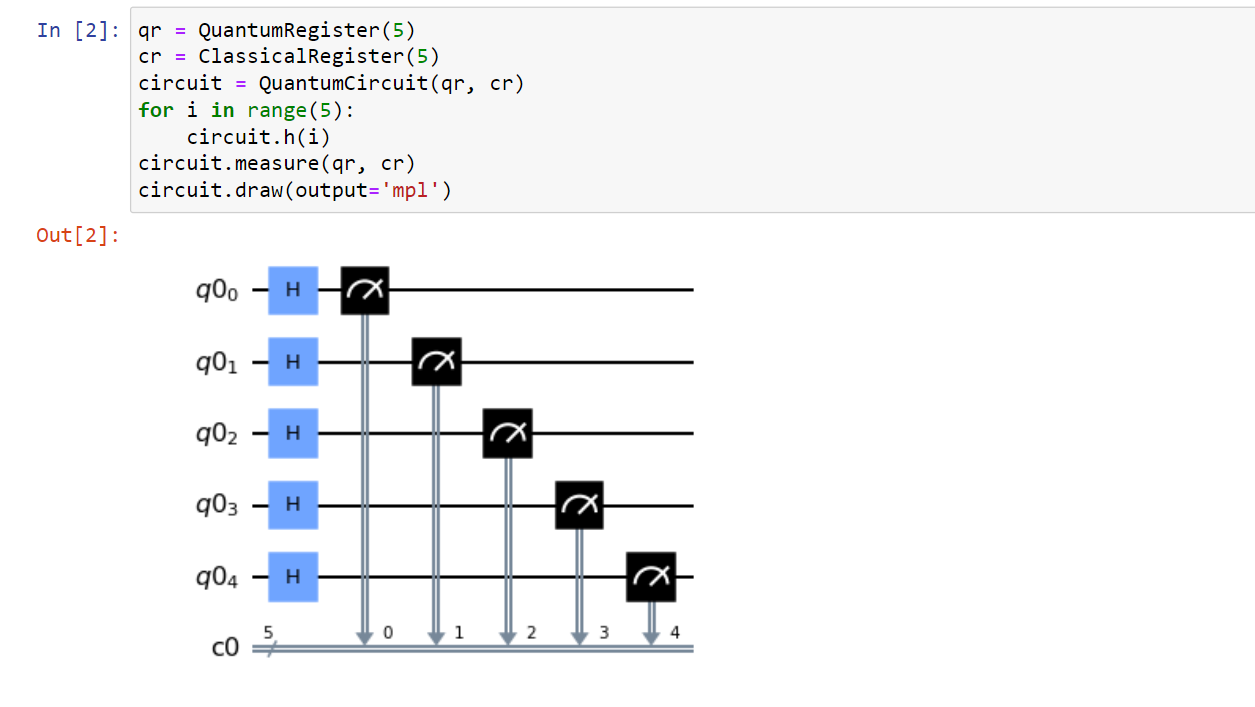}
    \caption{(Color online) Hadamard based quantum random number generator.}
    \label{fig:my_label}
\end{figure}

 Suppose we have 5 qubits, all in initial state $\ket{0}$ (as shown in Fig. \ref{fig:my_label}), then the application of Hadamard gate (on each qubit) will bring each qubit in the equal superposition state $\frac{\ket{0}+\ket{1}}{\sqrt{2}}$ and the combined state of the system will be $\left(\frac{\ket{0}+\ket{1}}{\sqrt{2}}\right)^{\otimes5}$. If we perform a measurement on any of the qubit then the superposition will collapse into any one of the state namely  $\ket{0}$ or $\ket{1}$ with equal probability. If the circuit shown in Fig \ref{fig:my_label} is run on a quantum computer, then each shot of the run will lead to generation of 5 bits of randomness.  IBM allows us to run $20000$ shots in a go, and a five qubit quantum computer is free, so one can easily generate $20000\times 5=100000$ bits of randomness in one run, and write the outcome sequentially using the command ``$result.get\_memory()$'' \cite{qiskit}. If one needs a longer string of randomness, one can run the circuit multiple times and concatenate the outcomes. This approach is already adopted in several works \cite{tamura2020quantum,tamura2021quantum,kumar2022quantum}.  This was the simplest possible approach, now we may discuss a slightly more involved approach involving entangled states.

\subsection{Entanglement based QRNG} 
\begin{figure}[h!]
    \centering
    \includegraphics[width = .9\textwidth]{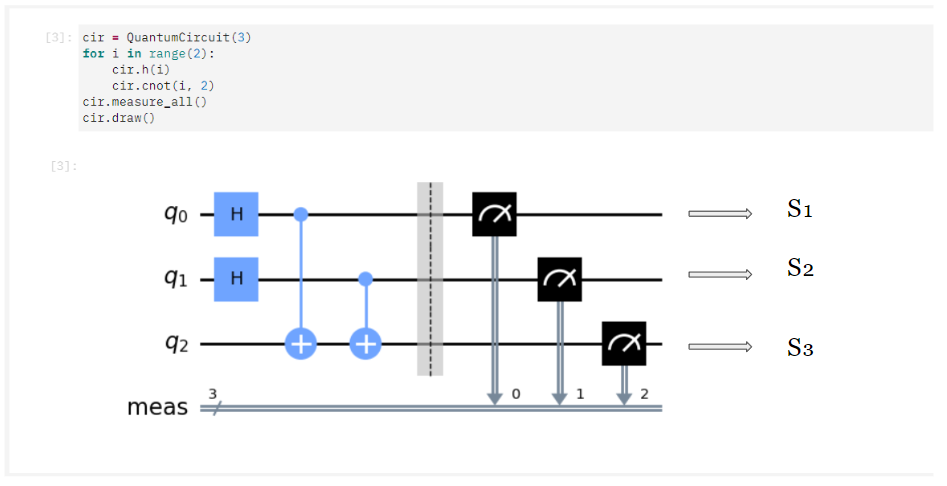}
    \caption{(Color online) Entanglement based quantum random number generator.}
    \label{fig:entanglment}
\end{figure}

Entanglement is a special property of quantum systems by which the joint state of two or more particles can not be expressed as tensor product of the states of individual particles. This results in the creation of correlations between particles that can not be explained by any classical theory. Such entangled qubits can be created by the application of Hadamard and CNOT gates. For example, see the circuit shown in  Fig. \ref{fig:entanglment}. This quantum circuit can be used to generate an entangled qubit $\ket{\Phi}= \frac{1}{2} \{ \ket{000}+\ket{011}+\ket{101}+\ket{110}\}$.   All the three qubits are  entangled in such a fashion that after collapse on measurement in computational basis, the XOR of any two output bits is equal to the third bit. When this circuit is run over a quantum computer, then the output result can be used to generate three sequences of random numbers.  Such an entanglement based QRNG has an added advantage that due to entanglement (correlation) the amount of entropy present in each sequence is the same. Hence, one can perform a  randomness test on just one of the sequences and the same amount of randomness will be guarantied to be present in other sequences, too. In general, one can use $'n'$ qubits to generate  $'n'$-qubit entangled state and hence produce $'n'$ sequences of random numbers. One sequence can be used to test for randomness while the rest can be used elsewhere. In fact, such a scheme was proposed by Jacak et al. \cite{jacak2020quantum} to not only generate random numbers but also allow a trusted third party to publicly perform tests on the generated random numbers.
Moving further, we want to emphasize that there is a difference between nonlocality and entanglement. Every nonlocal state is entangled, but the converse is not true for the mixed states and appropriate use of nonlocality enables us to generate certifiable randomness. Entanglement alone is not sufficient for the production of certifiable randomness. Though for pure states, all entangled states are nonlocal, to ensure self-testing of the source of randomness and thus to produce self-testing or semi-self-testing QRNGs, we will require to perform nonlocality tests.  In the next section, we will see how one can exploit the nonlocal properties of entangled states to manufacture devices which not only produces random numbers, but also tests whether the source of randomness is quantum mechanical.

 \subsection{Self-testing and semi-self-testing QRNGs} 

The quantum random number generators discussed till now are based on the trust that IBM is actually running those circuits on a quantum hardware and then sending us the output. So, the user has to trust the vendor that real quantum systems are the source of entropy rather than some simulations. The same is true for many commercially available QRNGs including Quantis, but  a user may like to be sure against the malefic interest of any adversary. So, self-testing or semi-self-testing QRNG is required. The idea behind the self-testing QRNG is to consider the device as a black box and then observe only the  
 input and output statistics \cite{colbeck2011private}. The certification of randomness is done by observing input-output statistics and matching them with the expected distribution produced by the laws governing the physical system used. In the case of QRNGs, it is done by observing the nonlocal behavior of the quantum physical systems such as by use of Bell tests \cite{dhara2013maximal}. The violation of Bell inequality confirms that the output bits are being generated by using some nonlocal physical system which can never be generated by any classical system as they are not nonlocal.  This idea of certifying the randomness by observing the nonlocal behavior of the device was first used in the field of quantum cryptography  \cite{selftest6} to develop {device independent quantum cryptography(DI-QKD)} \cite{DIcryp}, but now the field has grown beyond that and device independence is established in many contexts including the generation of device independent random numbers \cite{pironio2010random,giustina2013bell,liu2018device,liu2021device}. 
    
Self-testing devices are very strict with the testing of the randomness because they do not trust the device/manufacturer, but the downside of them is that they are very hard to implement. Further, they are also very slow as compared to the trusted devices which makes their use limited. However, if one is not concerned about the malicious intentions of the manufacturer and is only concerned about the noise or some particular part of the system then semi-self testing schemes can be adopted. Semi-self-testing devices slightly loosen the thread on security/credibility side, but increases the device performance. Semi-self-testing devices \cite{Semi_self} are intermediate between trusted and self-testing devices and depending on the part of system about which one is concerned,  adequate certification can be developed. For example, depending upon the certification of different parts of the devices one can have measurement-device independent QRNGs \cite{nie2016experimental} and source-device independent QRNGs \cite{cao2016source}. To the best of our knowledge, the idea of using cloud based quantum computers to generate certifiable self-testing random numbers remains an unexplored area of research. So, in the next section, we will propose a scheme for the generation of certifiable self-testing random numbers via violation of CHSH inequality and then use IBM quantum computer to implement the scheme.

\section{Quantum random number generation using violation of CHSH inequality} \label{CHSH_QRNG}

In this section, we would like to propose a method by which we can generate device independent quantum random numbers by use of IBM quantum computers that are available on the cloud. The random numbers generated by such a method can be tested via a form of CSHS inequality. In the following, we will first briefly introduce the quantum game version of CSHS inequality and then use that quantum game to generate high quality random numbers. 

\subsection{Description of CHSH-Game}
Let us suppose that we have two players Alice and Bob who are separated from each other. Consider, a third participant Charlie who is playing the role of a referee. Charlie provides Alice and Bob with a binary input ({0,1}) namely $x$,$y$ respectively  \cite{chshgame,chsh2}. Alice and Bob will then provide their respective binary outputs as $a$ and $b$. Alice and Bob will win the game if  $x\cdot y = a \oplus b$.
\begin{table}[h!]
\centering
		    \begin{tabular}{|p{2cm}|p{2cm}|p{3cm}|p{4cm}|}
			\hline
				\textbf{$(x,y)$} & \textbf{$(x.y)$} & \textbf{For win $a \oplus b$} & \textbf{$(a,b)$} \\
				\hline 
            $(0,0)$ & 0 & 0 & $(0,0)$ or $(1,1)$ \\
            \hline
             $(0,1)$ & 0 & 0 & $(0,0)$ or $(1,1)$ \\
            \hline
             $(1,0)$ & 0 & 0 & $(0,0)$ or $(1,1)$ \\
            \hline
             $(1,1)$ & 1 & 1 & $(0,1)$ or $(1,0)$ \\
            \hline
			\end{tabular}
        \caption{Overview of all possible settings to win the CHSH game}
        \label{table:CHSH}
	\end{table} 
 There are four possibilities in the game depending on values of $x$, $y$ as described in Table \ref{table:CHSH}. The best classical strategy for Alice and Bob to win the game is to provide the same bits as their outputs, i.e. their bits should be either $(0,0)$ or $(1,1)$. In this way, they can maximize their chances of winning. Further, it has been proved that  no classical strategy can help Alice and Bob  win with probability more than $\frac{3}{4}$ \cite{chsh2}.  However, if Alice and Bob use a quantum strategy to play the game, they can win with max probability of 85\%, thus the CSHS inequality  $P(ab|x.y=a \oplus b) \leq \frac{3}{4} $ can be violated in the quantum regime. The question is how to perform CSHS inequality tests in an IBM quantum computer and how can that be exploited to produce DI-QRNGs? The following subsection is aimed to address these questions.
\begin{figure}[h]
    \centering
    \includegraphics[width = .5\textwidth]{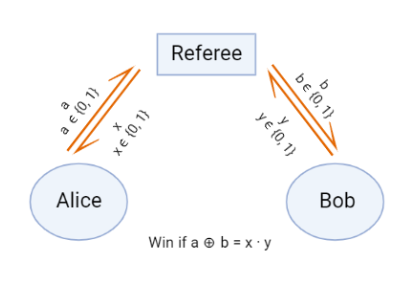}
    \caption{(Color online) Two players Alice and Bob playing CHSH game on their respective device. }
    \label{fig:chsh}
\end{figure}

\subsection{CHSH game on IBM's quantum computers: A self-testing mechanism for device-independent QRNG}

In this section, we will play the CHSH game on IBM quantum computers and use such a game to generate self-testing quantum random numbers. 

\subsubsection{Goal:}

The goal of the experiment is to observe the violation of CHSH inequality 
\begin{equation}
    P(ab|x.y=a \oplus b) \leq \frac{3}{4}
\end{equation}
  
\subsubsection{Game settings}
\begin{itemize}
    \item 5 quantum computers have been used. 
    \item Total 100 rounds of the game has been played. 
    \item Each round consists of 1000 shots.  
    \item Charlie's input $x$ and $y$ have been generated independently on two different quantum computers through Hadamard based QRNG.  
\end{itemize}

\subsubsection{Quantum devices used}
The tests have been performed on IBM quantum computers namely $ibmq\_quito$, $ibmq\_belem$, $ibmq\_manila$, $ibmq\_lima$ and $ibmq\_jakarta$. All the devices have 5 qubits except $ibmq\_jakarta$ which has 7 qubits.  $ibmq\_quito, ibmq\_belem$ and $ibmq\_manila$ were accessed for running the experiment on the date $23^{rd}$ May 2023 and $ibmq\_lima$ and $ibmq\_jakarta$ were accessed on $24^{th}$ May 2023. They differ in many parameters like T1 and T2 \cite{T1_T2}, quantum volume \cite{quantvol}, circuit operations per second (CLOP), frequency, median readout, and errors which has been detailed in Table \ref{table2}. Now, the idea is to treat the device as a black box and only consider the output statistic corresponding to the respective inputs.  Every device that has been used has a maximum limit of 100 circuits at a time which means that we can play 100 rounds of games at a time. Since each circuit is allowed to run for a maximum of 20000 shots, so in other words, each round/setting of the game can be played for the maximum of 20000 times. For our experiment, we have considered 1000 shots only for every game setting.
\begin{table}[h!]
\centering
\begin{tabular}{|l|c|c|c|c|l|}
\hline \multicolumn{1}{|c|}{ Device } & Qubits & Max. Circuit & Quantum Vol.  & T1, T2 ($\mu s$)  & Processor\\
\hline ibmq\_lima & 5 & 100 & 8 &98.68, 115.32 & Falcon $\mathrm{r} 4 \mathrm{~T} $ \\
\hline ibmq\_quito & 5 & 100 & 16 & 96.83, 104.39 & Falcon $\mathrm{r} 4 \mathrm{~T}$ \\
\hline ibmq\_belem & 5 & 100 & 16 & 101.42, 98.85 & Falcon $\mathrm{r} 4 \mathrm{~T}$ \\
\hline ibmq\_manila & 5 & 100 & 32 & 141.15, 56.53 & Falcon $\mathrm{r} 5.11 \mathrm{~L}$ \\
\hline ibmq\_jakarta & 7 & 100 & 16 &136.95, 38.99 & Falcon $\mathrm{r} 5.11 \mathrm{H}$ \\
\hline
\end{tabular}\caption{Parameters of different IBM quantum computers used in the experiment.}  \label{table2}
\end{table}

\subsubsection{Quantum strategy for CHSH game}

\begin{figure}[h]
    \centering
    \includegraphics[width = .8\textwidth]{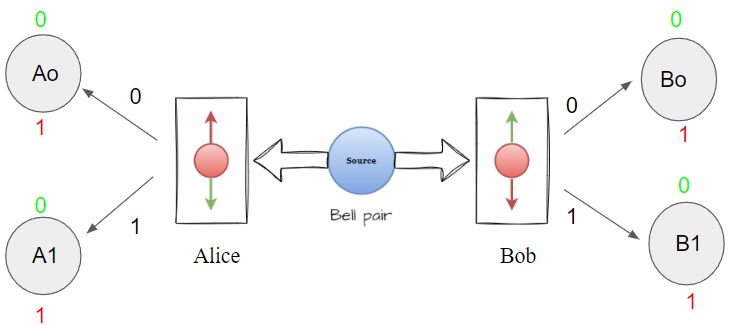}
    \caption{(Color online) Alice and Bob using quantum strategy in CHSH  game. }
    \label{fig:epr}
\end{figure}

Let us now describe the quantum strategy adopted by Alice and Bob to win the game with a probability greater than that allowed by the best classical strategy. Alice and Bob are provided some binary inputs  $x$,$y$ and they are not allowed to communicate with each other. Let us now consider that Alice and Bob share a Bell state\footnote{$\ket{\psi^{\pm}}=\frac{1}{\sqrt{2}}\{\ket{00}\pm\ket{11}\}, \ket{\phi^{\pm}}=\frac{1}{\sqrt{2}}\{\ket{01}\pm\ket{10}\}$}. One qubit of the Bell state is with Alice while the other one is with Bob. The strategy of Alice and Bob as shown in Fig. \ref{fig:epr} is as follows. Alice and Bob respectively receive input $x$,$y$ from Charlie and based on the input received from Charlie, they use the  following basis to measure the state of their qubit. 
\begin{itemize}
    \item if $x = 0$, Alice measures in $A_o$
    \item if $x = 1$, Alice measures in $A_1$
    \item if $y = 0$, Bob measures in $B_o$
    \item if $y = 1$, Bob measures in $B_1$    
\end{itemize}

The measurement basis $A_o$ is the horizontal axis and $A_1$ is $\pi/4$ angle rotated axis from $A_o$. Similarly, Bob's measurement basis $B_o$ is $\pi/8$ degree rotated axis from $A_0$ and $B_1$ is $-\pi/8$ angle rotated axis from the same reference. We know that the measurement of a qubit state is mathematically equivalent to a collapse of the state vector, hence when we are measuring a qubit in two different bases then the probability with which the state collapses and produces the same output in both the measuring bases is given by $\cos^2({\theta})$ where $\theta$ is the angle between both the bases.
\begin{figure}[h]
    \centering
    \includegraphics[width = .7\textwidth]{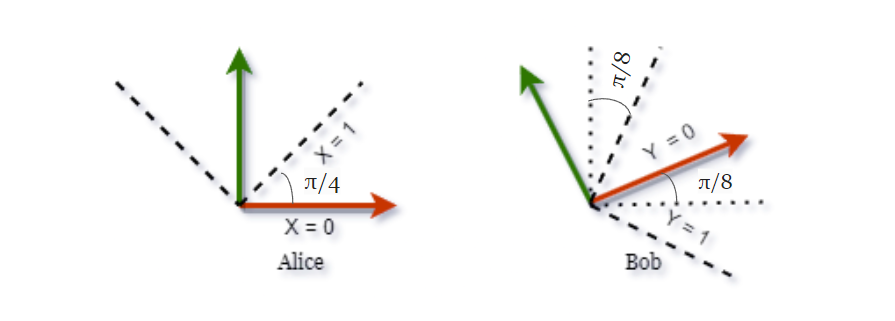}
    \caption{(Color online) Measurement bases of Alice and Bob. }
    \label{fig:bases}
\end{figure}
\begin{figure}[h]
    \centering
    \includegraphics[width = .5\textwidth]{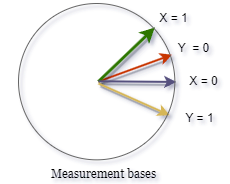}
    \caption{(Color online) Planer statevector representation of measurement bases.}
    \label{fig:measurementl}
\end{figure}

We may now elaborate on one round of the game for a better understanding of the strategy. Consider the case when inputs are $(x = y = 0)$. Alice measures her qubit in $A_o$ basis and Bob measures his qubit in $B_o$. As we can see in the Fig. \ref{fig:bases} and \ref{fig:measurementl}, the angle between both the bases is $\pi/8$, so the probability to get same results is given by $P(a=0, b=0) = P(a=1, b=1) = \cos^2(\pi/8)$.
 In this case $x.y = 0$, hence $a \oplus b = 0$ is the winning condition which can happen only if both $a$ and $b$ are $0$ or both  $a$ and $b$ are $1$. Consequently, the probability of winning the game for  Alice and Bob is $P_{win}(x = 0, y = 0 ) = \cos^2(\pi/8) = 0.85$. In fact, this is more than that allowed by use of only classical resources. Similarly for all other cases as elaborated in Table \ref{table:CHSH}, we can see that if Alice and Bob make measurements in the appropriate basis, then the probability of winning the game for Alice and Bob is $0.85$. So, we can conclude that if Alice and Bob have Bell states with them, then they can win the game with a probability which can never be achieved by use of classical strategy.

\subsubsection{Quantum circuit for the game and generation of  quantum random numbers}

\begin{figure}[h!]
    \centering
    \includegraphics[width = .8\textwidth]{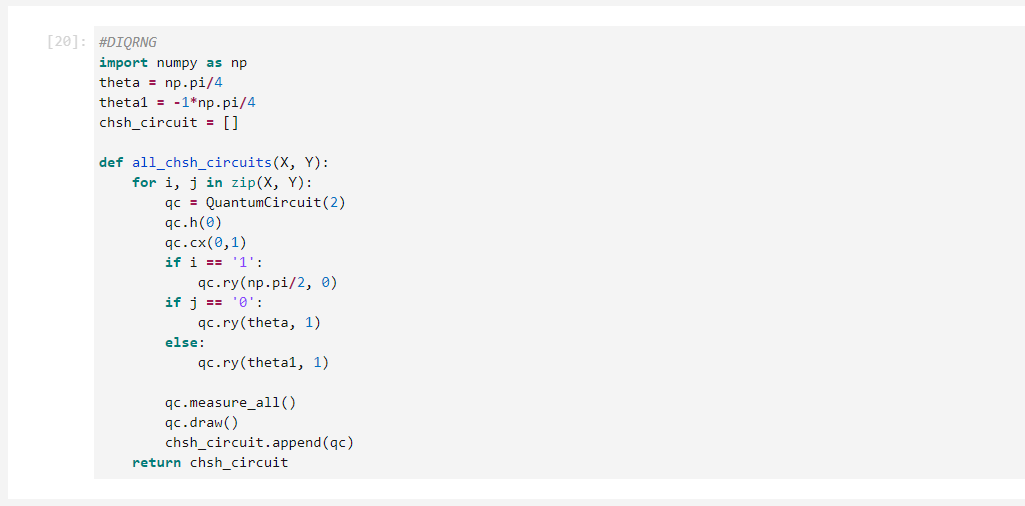}
    \caption{ Evolution of Bell states for the measurement in the required bases. }
    \label{fig:evolution}
\end{figure}

In order to play the quantum game on IBM quantum computer, first we generate a Bell state with first qubit residing with Alice while the second qubit is with Bob. Now, Charlie provides random inputs to Alice and Bob based on which they choose the basis to measure their qubits. The random bits sent by Charlie to Alice and Bob are simulated by storage of quantum random numbers generated by a Hadamard based QRNG in which we have used different quantum computers to generate the sequence. In this way, Alice and Bob also receive independently generated quantum random numbers to decide upon their basis for measurement of their qubits. 
\begin{figure}[h!]
  \centering
  \begin{minipage}[b]{0.4\textwidth}
    \includegraphics[width=\textwidth]{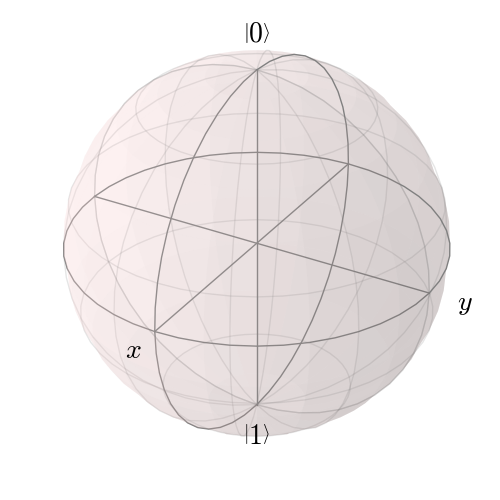}
    \caption{(Color online) 3-D Bloch sphere.}
    \label{fig:bloch}
  \end{minipage}
  \hfill
  \begin{minipage}[b]{0.4\textwidth}
    \includegraphics[width=\textwidth]{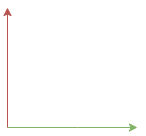}
        \caption{(Color online) Statevector representation system of measurement bases.}
        \label{fig:planar}
  \end{minipage}
\end{figure}
Now, since the IBM quantum computer does not provide any way to measure in any arbitrary basis. So, the idea was to rotate the state by same angle instead of rotating the basis. In this way, the angle between the state and the basis of measurement remains the same and hence the statistics of the measurement will be the same. An  important thing to note is that whenever we wanted the state to be rotated by $\pi/4$, $\pi/8$ or $-\pi/8$ we applied the $R_y(\theta)$ gate for $\theta = \pi/2, \pi/4$ and $-\pi/4$, respectively because $R_y(\theta)$ gate rotates the state around Y-axis by an angle $\theta$. We can see in Fig. \ref{fig:bloch} of Bloch sphere, state $\ket{0}$ is vertically upward and $\ket{1}$ is vertically down. The orthogonal states are actually at 180 degrees from each other. However, in the planer state vector representation (see Fig. \ref{fig:planar}), we can see that state $\ket{0}$ is at horizontal and $\ket{1}$ is at vertical, i.e., $90^{\circ}$ from each other. In general, the radian space of $\theta$ in the Bloch sphere has been compressed in the radian space of $\theta/2$ in the planer form. So if we want to rotate the state by $\theta$ degree in planer form, then we need to rotate it by $2\theta$ in the Bloch sphere because all the rotations operators such as $R_x$, $R_y$, $R_z$  rotate the state vector by the given amount around their respective axis in the reference to Bloch sphere. For our experiment, the program for rotation of qubits by Alice and Bob is shown in Fig. \ref{fig:evolution}.

\subsubsection{Calculation of the winning probability for each round of the game}

\begin{figure}[h!]
    \centering
    \includegraphics[width = 0.8\textwidth]{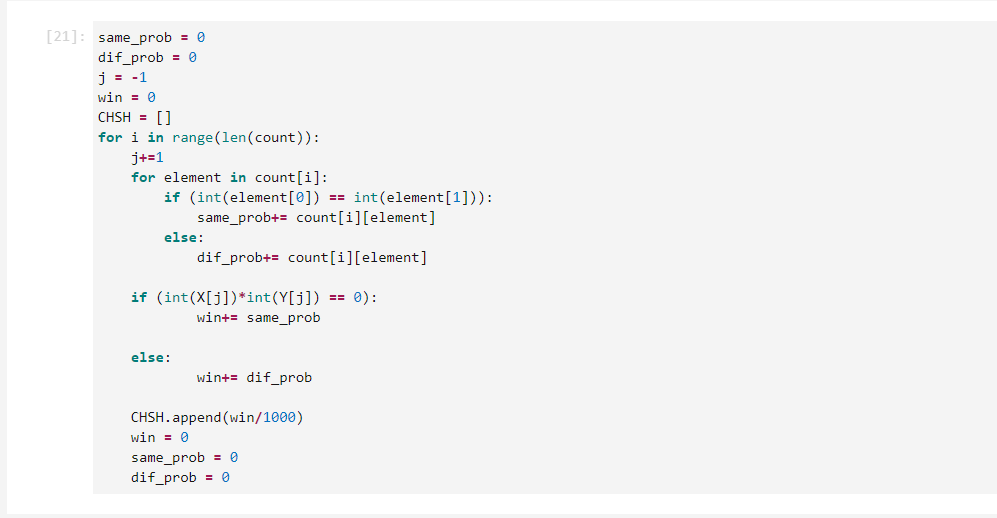}
    \caption{Calculation of the winning probability after executing the circuit on a quantum device.}
    \label{fig:P_win calculation}
\end{figure}

\begin{figure}[h!]
    \centering
    \includegraphics[width = 0.65\textwidth]{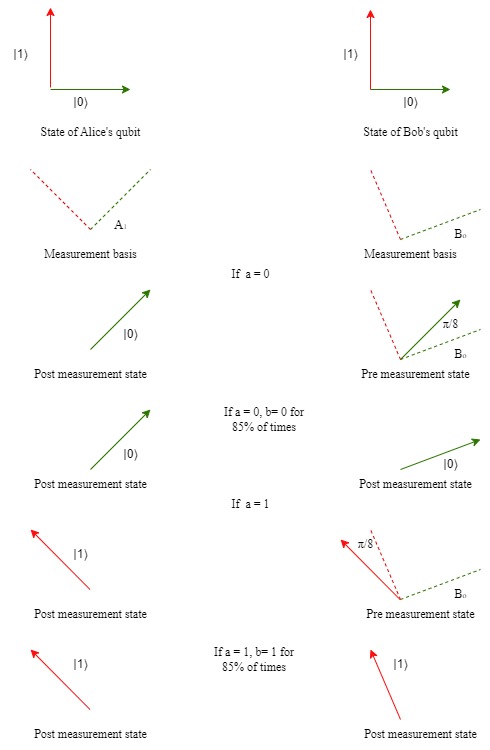}
    \caption{(Color online) Pictorial description of one round of the CHSH game}
    \label{fig:chsh-example}
\end{figure}

 Consider the case $(x = 1, y = 0)$. Alice needs to measure her qubit in $A_1$ and Bob in $B_0$. Alice applies $R_y$($\theta = \pi/2$) on her part of Bell state and Bob applies $R_y$($\theta = \pi/4$) on his part as it will rotate the state by $\pi/4$ degree and $\pi/8$ degree, respectively in the state vector representation. In order to win, Alice and Bob need to have the same result ($a = 0, b = 0$) or ($a = 1, b = 1$). As we can see no matter whether the output of Alice is 0 or 1, if we are using the quantum strategy described earlier, then Bob's output will be correlated to that of Alice such that they have the same result with the probability $\cos^2(\pi/8) = 0.85$. Similarly, for the other cases too, the game is won by Alice and Bob with probability $0.85$. In the execution of the circuits, the results of each round of the game is stored in the dictionary called \textbf{\textit{'counts'}}. The quantum circuit for each round of the game is executed 1000 times. As soon as a circuit gets executed we store the fraction of time when $a = b$ out of the 1000 cases and store it in the variable \textbf{'same\_prob'} and the fraction of times $a \ne b$ is stored in the variable \textbf{'dif\_prob'}. We then check for the $x \cdot y $ and if it is 0 then the value of 'same\_prob' is stored in \textbf{'win'} or else if it 1 then 'dif\_prob' is stored. Finally, the value of 'win' is stored for each round in \textbf{'CHSH'} as shown in Fig. \ref{fig:P_win calculation}. Further, Fig. \ref{fig:chsh-example} pictorially describes one round of CHSH game.

\subsection{Generalized quantum strategy for winning CHSH game}
The quantum strategy used in the CHSH game has several variations of it with different winning probabilities. In our case, we have used the basis $A_0$, $A_1$, $B_0$, and $B_1$ that are $\pi/4$, $\pi/8$, and $-\pi/8$ rotated from $A_0$ respectively. Where $A_0$ is the horizontal axis on the planer Bloch sphere. One important thing to observe is that until the relative arrangement of the bases is in the order shown in Fig. \ref{fig:bases}, no matter where they lie on the planer Bloch sphere, their output results would give similar statistics. In general, $A_0$ can be rotated at any angle from the horizontal as long as the relative position of other bases is fixed with respect to Bob.  The probability of a respective outcome depends solely on the angle between the state and basis. As long as the relative positioning of the basis is such that the angle between ($A_0$, $B_0$), ($B_0$, $A_1$), and ($A_0$, $B_1$) bases are $\pi/8$ and the angle between ($A_1$, $B_1$) bases is $3\pi/8$ for the cases where $x\cdot y = 0$ both the players will get the same result with maximum winning probability of $\cos^2({\pi/8}) = 0.85$. For the cases were $x\cdot y = 1$ both will get same bit with probability $\cos^2({3\pi/8}) = 0.15$. Since, for such cases, the game requires both the players have different bits, probability of winning will be $1 - \cos^2(3\pi/8) = 0.85$. Hence no matter how much angle $A_0$ makes with horizontal, as long as their relative positions are defined, states will always collapse on the basis with the same statistics.

\subsection{Results}

\begin{figure}[h!]
    \centering
    \includegraphics[width = .70\textwidth]{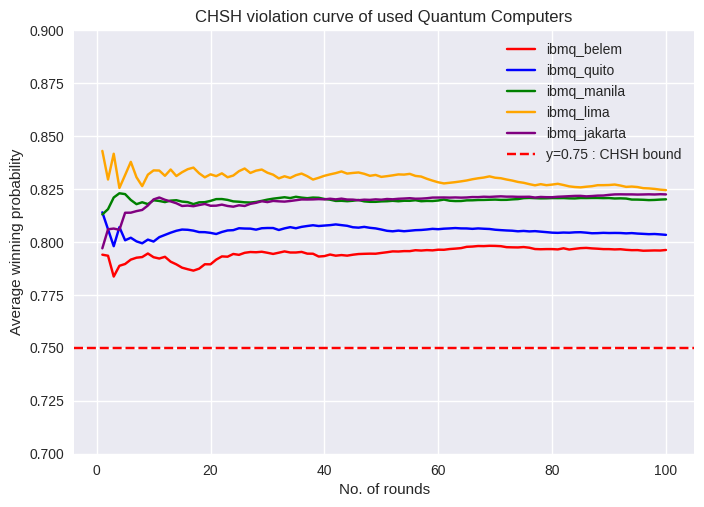}
    \caption{(Color online) Plot of the average winning probability versus number of rounds for different IBM quantum devices.}
    \label{fig:Pwin1}
\end{figure}

\begin{figure}[h!]
    \centering
    \includegraphics[width = .70\textwidth]{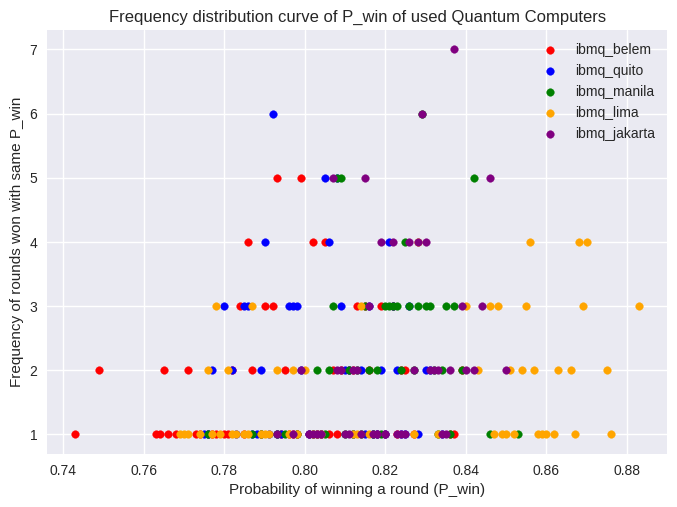}
    \caption{(Color online) Plot of the frequency of rounds won versus the probability $P_{win}$ for different IBM quantum devices.}
    \label{fig:Pwin2}
\end{figure}

\begin{figure}[h!]
    \centering
    \includegraphics[width = .70\textwidth]{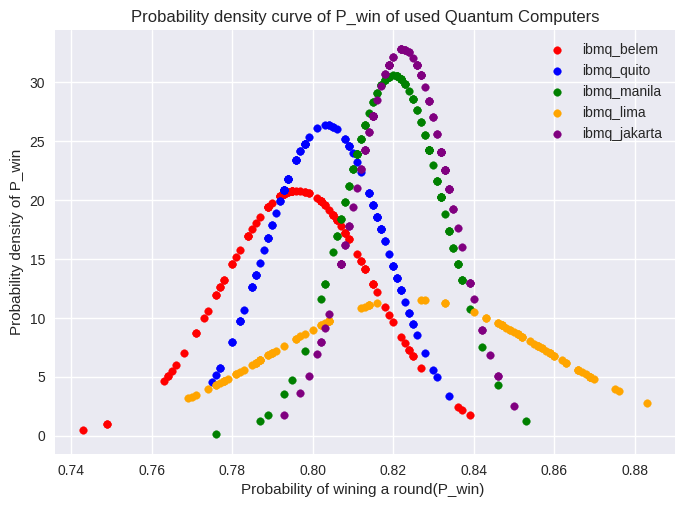}
    \caption{(Color online) Plot of the probability density of winning rounds versus the probability $P_{win}$ for different IBM quantum devices.}
    \label{fig:Pwin3}
\end{figure}

We have experimentally realized the CHSH game on different IBM quantum computers and results have been tabulated in Table \ref{table:result}, where we can clearly see that for each of the  IBM quantum computers used, the average winning probability for each case is greater than 0.75. This clearly shows that the underlying phenomena is nonlocal in nature. 
\begin{table}[h!]
\centering
\begin{tabular}{|p{0.15\linewidth}|p{0.15\linewidth}|p{0.13\linewidth}|p{0.13\linewidth}|p{0.13\linewidth}|p{0.15\linewidth}|}
\hline Device & Shots & Minimum winning probability & Average winning probability &  Maximum winning probability & Standard deviation$(\sigma)$   \\
\hline ibmq\_belem & 1000 & 0.743 & 0.79622 & 0.839 & 0.0191 \\
\hline ibmq\_quito & 1000 & 0.775 & 0.80335 & 0.834 & 0.0151 \\
\hline ibmq\_manila & 1000 & 0.776  & 0.82014 & 0.853 & 0.0130 \\
\hline ibmq\_lima & 1000 & 0.769 & 0.82448 & 0.883 & 0.0344 \\
\hline ibmq\_jakarta & 1000 & 0.793 & 0.82245 & 0.850 & 0.0121 \\
\hline
\end{tabular}\caption{Tabulation of performance statistics of used IBM quantum computers for the CHSH game. }
 \label{table:result}
\end{table}
Further, results of the CHSH game played on available quantum computers has been summarized in Figures \ref{fig:Pwin1}, \ref{fig:Pwin2} and \ref{fig:Pwin3}. From all the observations, we can see that since the winning average probability is more than that allowed by the best allowed classical strategy, so the random numbers that are being generated at Alice and Bob's end have quantum origin and its is being certified by violation of CHSH inequality.

Since in CHSH game played on IBM quantum computers, we can observe that Alice and Bob are winning with the probability that is always higher than that maximally allowed by the use of only classical resources. So, we can clearly see that some nonlocal property of quantum states is being used by Alice and Bob to generate correlated random numbers. Hence one can be sure that random numbers generated by such a method are actually generated by a quantum source. Hence, such a mechanism can be used for certifications of the random numbers that have been generated. Now, we can use either of the sequences of random numbers generated at Alice's end or Bob's end to produce a device-independent random number. We stored the sequence of random numbers generated by such a process and then performed NIST test too and as expected they passed the tests. 

Next, we would look to investigate the possibility of loopholes (if any) being present during the course of the experiment.

\section{Loopholes and their closings for device independent QRNG} \label{loopholes}
Some assumptions are considered before performing any experiment for testing of the Bell's inequality or any associated form of it like CHSH inequality. These loopholes need to be closed by scientists during the experiments to make sure that results are indeed of quantum origin.  In the following, we will discuss each loophole one by one and the methods we applied to close them. 

\subsection{Freedom of choice $x, y$ } 
This states that the measurement settings of one player should not be known to another player. Both the measurement outputs $a$ and $b$ should be independent of other player's input or else the players can use this information to produce a biased output. They need not measure the entangled qubits, as in that case, they can prepare the outputs that would satisfy the criteria. The $x$ and $y$ should also be randomly generated for the fair distribution of each game round \cite{freechoiloophole}. There should be no inter/intra correlation between the elements of the sets $x$ and $y$.  In order to keep $x$ and $y$ uncorrelated they are generated randomly and independently using different quantum computers. Quantum computer is unaware about the circuits before they are fed at the time of execution. Once the time for execution comes, the quantum computer is fed with the circuits for the execution. Before that, the quantum computer doesn't know about the experiment and measurement settings. Each round of the game is designed in a separate circuit with separate measurement settings. Each execution of the circuit is done independently, and the average probability has been calculated to observe the variations in the statistics of  violations. So, we can see that this loophole is closed as $x$ and $y$ have been generated independently with no correlations between them.

\subsection{Fair sampling or detection loophole}
If one chooses a small sample from an ensemble, then it might not behave statistically as the entire ensemble will behave. Similarly, from a small group if someone concludes that the whole ensemble behaves in the same way then it amounts to an abuse of the 'fair sampling assumption \cite{fair}'. The observation of violation in a small part of the experiment cannot conclude that the whole experiment violates the CHSH inequality. There are multiple physical hardware realizations of the Bell experiment. For example, if we are using entangled photons for the Bell experiment we might lose some photons, as photons are easily lost due to the correlations of the photons with the environment, over time and low device efficiency \cite{lowdevice}. Then one should have a large sample to draw fair conclusions. Similarly, for our case of entangled superconducting qubits, our inferences should be drawn based on a statistically large number of experimental runs.  

In order to close the fair sampling loophole, each round is played several times. For standard readings, 1000 shots of each round are executed. In this way, we don't report  whether a round is won or not by just a single shot of experiment, rather for each round the game is played for 1000 shots. In this way, errors if any are statistically eliminated to report our observations regarding the win or loss for a particular round. In the experiment, we played a round 1000 times, then we calculated based on input statistics that out of 1000 shots how many times both the players produced a required bit i.e., we report with what probability each round of the game is won.

In this way, by increasing the number of shots we have increased the authenticity of our output statistics. Also, all the devices which we have used in the experiment are based on the superconducting qubit model. Further, the superconducting qubits are highly efficient. The possibility of getting an erroneous measurement result at Alice and Bob's end is very very low. Due to this very small error bar, the violation of CHSH inequality as reported in the experiment remains valid. So, we can be pretty sure that the experiment is free from this loophole.

\subsection{Locality loophole}

Locality loophole is in fact the most difficult loophole to be closed as here one has to make sure that Alice and Bob are not communicating locally. So, when any  round of the game starts, no communication is allowed between the parties \cite{locloop}. Information traveling even at the speed of light from one of the devices should not be allowed to reach the second device before both the devices have made their respective measurements. The locality loophole is one of the loopholes that can not be closed on the IBM quantum computers. The quantum computers of IBM use superconducting qubits. In these types of hardware models, qubits are connected to different qubits in different arrangements on a very compact atomic scale. There is no way we can stop the information from one qubit to reach the other qubit without having access to the hardware part of the computers. The different architecture of the device is what differentiates the fundamental gates of different quantum computers. The locality loophole can be closed by properly separating the players so that there is no possibility of communication between the players. This loophole has been closed in a few physical experiments by properly setting the time of measurement and then the random orientation of the device \cite{Weihs_1998, Erven_2014}. In Ref. \cite{locloop} authors have managed to do this by decreasing the time duration between the measurement of both devices and separating both devices, such that not even light can reach from one device to another within the time. But for our case since, the superconducting qubits are not space like separated, so we have not been able to close this loophole. In fact, locality loophole is indeed very difficult to close for IBM quantum computers.

\subsection{Memory loophole}
The memory loophole considers the assumption that the device used in the experiment must not be used twice. It entertains the fact if the device is storing the results of the previous experiments then it can fake the results to produce the required results. The results of each round can be automatically deleted in Qiskit if we choose the \textbf{\textit{'memory' }} parameter to be 'False'. This will delete the outputs of the previous execution. But still, the device is the same. We cannot use another device for the next round, it would become very inefficient to wait in the queue for the execution of the next round of the game. Instead, we consider this loophole to be still open, but in principle, this loophole can be closed if we obtain dedicated access to the IBM quantum computers.

\subsection{Superdeternism}
We know that the whole universe is evolving under the laws of physics. So if we know all the laws of physics and make a generalized differential equation of all the laws then we can predict what's going to happen next. Or at least it was believed to be so by many physicists like Newton and Einstein himself. It makes the universe deterministic, but quantum mechanics is not deterministic in nature and is the best theory known so far that can beautifully describe many phenomena of nature. Quantum mechanics is now almost one century old, but people in physics still discuss the foundational questions and ask whether there is  free will or not. Because in the case of a deterministic universe, there is no need to randomly generate $x$, and $y$ as they will not be random at all, they will be deterministic. Superdeternisim \cite{freewill} is considered untestable till now. Generally, scientists consider it to be closed, because if this loophole is not considered close then there is no need for any other loophole to be closed, The experiment itself has become meaningless in a deterministic universe.  

To conclude we have been able to fully close certain loopholes such as freedom of choice and fair sampling loophole but with IBM quantum computers one can never close the locality loophole as qubits are within talking distance of each other. Further, the memory loophole is partially closed as the results of one experiment are automatically deleted before the start of the next experiment.  

\section{Conclusions} \label{conclusions}

We all know that the generation of random numbers is extremely important for various applications in many fields. The use of PRNG is not desirable for many sensitive applications such as cryptography. Quantum systems due to their intrinsic randomness seem beneficial for the generation of random numbers of very high quality, but even if one uses QRNG, one has to make sure that indeed the random numbers generated are of quantum origin. So the main motivation of this work is to introduce the use of CHSH game as a self-testing mechanism for the development of device independent quantum random numbers  (DI-QRNG) using IBM's quantum computer. So, in this work, we have proposed a scheme via which one can produce DI-QRNG using the quantum computers that are available on cloud. Since IBM quantum computers are available free of cost, so the generation of self-testing random numbers using the violation of CHSH inequality on IBM quantum computers is a very cheap and extremely reliable method. Further, staying within our limitations we have tried to close as many loopholes as possible. We hope that this study will help in providing new directions for the development of self-testing and semi-self-testing random devices via the use of quantum computers. 

 \section*{Acknowledgement} AY thanks Jaypee Institute of Information Technology, Noida for the support provided to perform this work. AP acknowledges the support from the QUEST scheme of Interdisciplinary Cyber Physical Systems (ICPS) program of the Department of Science and Technology (DST), India (GrantNo.: DST/ICPS/QuST/Theme-1/2019/14 (Q80)).

\section*{Data Availability} Experimental data will be made available on request.

\section*{Conflict of interest} The authors state that they do not have any competing interests. 

\bibliographystyle{ieeetr}
\bibliography{main}
\end{document}